\documentclass[prb,twocolumn,showpacs,superscriptaddress]{revtex4}
\usepackage{graphicx}
\usepackage{amsmath}
\usepackage[T1]{fontenc}
\date{\today}
\begin{document}
\title {Magnetic interactions in EuTe epitaxial layers and
EuTe/PbTe superlattices}
\author{H. K\k{e}pa}
\email{Henryk.Kepa@fuw.edu.pl}
\affiliation{Institute of Experimental Physics, Warsaw University,
Ho\.za 69, 00-681 Warszawa, Poland}
\affiliation{Physics Department, Oregon State University, Corvallis,
OR 97331, USA}
\author{G. Springholz}
\affiliation{Institut f\"ur Halbleiterphysik, Johannes Kepler University,
A-4040 Linz, Austria}
\author{T. M. Giebultowicz}
\affiliation{Physics Department, Oregon State University, Corvallis,
OR 97331, USA}
\author{K. I. Goldman}
\affiliation{Physics Department, Oregon State University, Corvallis,
OR 97331, USA}
\author{C. F. Majkrzak}
\affiliation{National Institute of Standards and Technology, Center for
Neutron Research, Gaithersburg, MD 20899, USA}
\author{P. Kacman}
\affiliation{Institute of Physics, Polish Academy of
Sciences, Al.~Lotnik\'ow 32/46, 02-668 Warszawa, Poland}
\author{J. Blinowski}
\altaffiliation[]{deceased}
\affiliation{Institute of Theoretical Physics, Warsaw University,
Ho\.za 69, 00-681 Warszawa, Poland}
\author{S. Holl}
\affiliation{Institut f\"ur Halbleiterphysik, Johannes Kepler
University, A-4040 Linz, Austria}
\author{H. Krenn}
\altaffiliation[Present address: ]{Institut f\"ur Experimentalphysik, Karl 
Franzens Universi\"at, A-8010 Graz, Austria}
\affiliation{Institut f\"ur Halbleiterphysik, Johannes Kepler University,
A-4040 Linz, Austria}
\author{G. Bauer}
\affiliation{Institut f\"ur Halbleiterphysik, Johannes Kepler University,
A-4040 Linz, Austria}

\begin{abstract}
The magnetic properties of antiferromagnetic (AFM) EuTe
epitaxial layers and short period EuTe/PbTe superlattices (SLs), grown 
by molecular beam epitaxy on (111) BaF$_2$ substrates, were studied by
magnetization and neutron diffraction measurements. 
Considerable changes of the N\'eel temperature as a function of the
EuTe layer thickness as well as of the strain state were found.
A mean field model, taking into account the variation of the exchange 
constants with the strain-induced lattice distortions, and the nearest neighbor
environment of a Eu atoms, was developed to explain the observed $T_{\text N}$
changes in wide range of samples.
Pronounced interlayer magnetic correlations have been revealed by neutron 
diffraction in EuTe/PbTe SLs with PbTe spacer thickness up to 60 \AA.  
The observed diffraction spectra were analyzed, in a kinematical
approximation, assuming partial interlayer correlations
characterized by an appropriate correlation parameter.
The formation of interlayer correlations between the
AFM EuTe layers across the  nonmagnetic PbTe spacer was
explained within a framework of a tight-binding model.
In this model, the interlayer coupling stems from the dependence of the total 
electronic energy of the EuTe/PbTe SL on the spin configurations in 
adjacent EuTe layers.
The influence of the EuTe and PbTe layer thickness fluctuations, 
inherent in the epitaxial growth process, on magnetic properties and 
interlayer coupling is discussed.
\end{abstract}

\maketitle
\section{Introduction}
\label{intro} In the last decade magnetic multilayer systems and
the giant magnetoresistance resulting from interlayer couplings
have been receiving considerable interest in both
applied and fundamental scientific research. Interlayer exchange couplings
in multilayers and superlattices (SLs) have been observed in a
large variety of structures composed of metallic ferromagnetic
(FM) layers alternating with nonmagnetic metallic
\cite{{gruen},{parkin}}, as well as nonmetallic 
\cite{{tosca},{briner}} spacer layers. These observations have stimulated 
extensive
theoretical studies that have resulted in a number of different models
for the mechanism of interlayer coupling such as the RKKY model,
the free--electron model, and several others. The most complete
theory unifying all previous approaches has been devised by Bruno
\cite{bruno}. However, neither interlayer coupling in systems
composed of two {\it nonmetallic} materials, nor mechanisms that
might give rise to coupling between {\it antiferromagnetic} (AFM)
layers have been considered in these works.

Yet, neutron diffraction data for three different SL systems
composed of AFM and nonmagnetic semiconducting materials,
reported in the mid-nineties \cite{{giebult},{nunez},{rhyne},{kepa02},{goldi}},
revealed the existence of pronounced interlayer correlations between 
the AFM blocks.
Also recently, coupling between FM--semiconductor layers has been found 
in EuS/PbS SLs \cite{kepa_epl}.
In these all-semiconductor systems,  the carrier concentration
is far too low to support any significant RKKY
interactions; in addition, the AFM layers do not
carry a net magnetic moment. Thus, the two main ingredients which were
believed to play a crucial role in interlayer coupling -- mobile
carriers and layer magnetization -- are absent in these cases. These results  
have clearly demonstrated that the magnetic interlayer coupling is not
restricted to structures  containing FM metallic components.
The proper understanding of correlations between the AFM semiconductor layers
may shed new light on the interlayer coupling mechanisms.

In this paper we present our systematic, experimental and
theoretical, studies of EuTe epitaxial films and short period
[(EuTe)$_m|$(PbTe)$_n$]$_N$ superlattices.
In Section \ref{materials} we describe the basic properties of the
constituent materials, sample preparation and the experimental
techniques used. The effects of the finite thickness and the
strain on magnetic properties of individual layers are treated
in Section \ref{strain}. Section \ref{coupling} is devoted to the
interlayer coupling determined by neutron diffraction
measurements. In its first three subsections, the neutron data
for a series of samples studied in zero, intermediate and high
magnetic in-plain field, as well as the effect of cooling in
external low magnetic field are presented. In the last subsection
we discuss the results of a theoretical model for the interlayer
coupling in a perfect AFM/nonmagnetic semiconductor SL and 
we compare the experimental results with the model predictions.

\section{Constituent materials, sample preparation and experimental techniques}
\label{materials}

\begin{figure}
\includegraphics[width=3in]{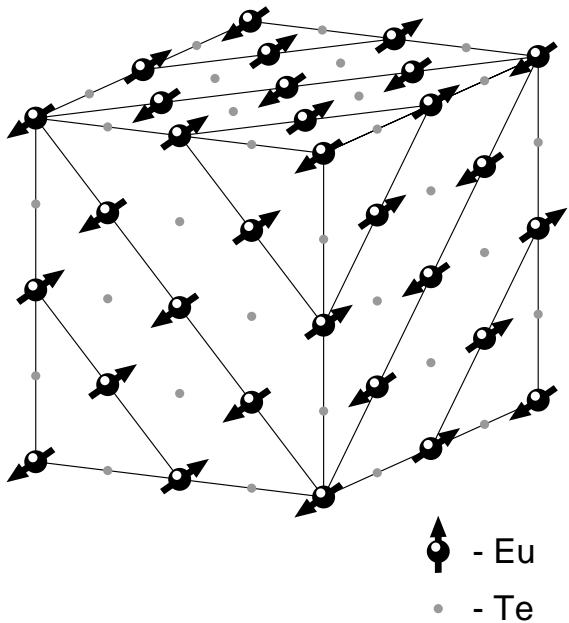}
\caption{\label{eute-cryst} Chemical and magnetic unit cell of EuTe with its
type II AFM structure consisting of ferromagnetically ordered (111) planes and
antiferromagnetic spin sequences along the [111] directions.}
\end{figure}

Bulk EuTe is a classical Heisenberg antiferromagnet with a N\'eel
temperature $T_{\text N}$ = 9.6 K \cite{wachter}. It crystallizes
in the NaCl structure with $a= 6.598$ \AA. The Eu$^{2+}$ ions with
$S=7/2$ and $L=0$ form an FCC spin lattice with dominant AFM
next-nearest neighbor interactions, and weaker FM interaction
between the nearest neighbors ($J_2/k_{\text B} = -0.23$ K,
$J_1/k_{\rm B} = 0.11$ K, respectively \cite{hihara}). Such a
$J_i$ combination leads to the Type II AFM ordering below
$T_{\text N}$, in which the spins are ferromagnetically ordered in
the (111) lattice planes, and the neighboring planes are coupled
antiferromagnetically to one another \cite{will} (see Fig.~\ref{eute-cryst}).
EuTe is a wide gap ($\sim 2.5$ eV) semiconductor with the $4f$
levels situated about  2 eV below the conduction band edge
\cite{wachter}. The diamagnetic constituent, PbTe, is a narrow gap
($\sim 0.19$ eV)  semiconductor, which also crystallizes in the
NaCl structure and has a bulk lattice constant of 6.462 \AA. This
yields a close lattice-matching to EuTe, with a mismatch of only
2.1 \% in the lattice constants.

The EuTe/PbTe SL samples were grown by molecular beam epitaxy on
(111) oriented BaF$_2$ substrates as described in detail in Ref.
\onlinecite{springholz}. Different
[(EuTe)$_m|$(PbTe)$_n]_N$ SL stacks with $m$ (varying from 2 to 10)
monolayers of EuTe alternating with $n$ (from 5 to 30) monolayers of PbTe,
were deposited on 1 -- 3 $\mu$m PbTe buffer layers.
In the paper the EuTe/PbTe SL with $m$
monolayers of EuTe and $n$ monolayers of PbTe in the SL period (bilayer)
is often denoted by an abbreviated symbol $(m/n)$.
The thickness of one monolayer is 3.81 \AA\ for EuTe
and 3.73 \AA\ for PbTe. To obtain sufficient neutron scattering
intensity, the number of periods $N$ was several hundred in all
cases. The layer thicknesses determined by high-resolution x-ray
diffraction agreed with the nominal thicknesses within $\pm 0.5$
monolayer. The electron concentration in the PbTe layers was
$\sim$ 10$^{17}$cm$^{-3}$, i.e., many orders of magnitude lower
than in metals, and the EuTe layers were semi-insulating.

The neutron experiments were performed at the NIST Center for Neutron 
Research. The instruments used were BT-2 and BT-9 triple-axis spectrometers
set to elastic diffraction mode, with a pyrolitic graphite (PG) monochromator
and analyzer, and a 5 cm PG filter in the incident beam. 
The wavelength used was $\lambda=2.35$~\AA~and  the angular collimation was
40 minutes of arc throughout. Additionally, a number of diffraction
experiments were carried out on the NG-1 reflectometer operated at neutron
wavelength $\lambda = 4.75$~\AA. The latter instrument yielded a high 
intensity, high resolution spectra with a negligible instrumental broadening 
of the SL diffraction lines. All the magnetic diffraction spectra reported
here have been measured at 4.3 K.
\begin{figure}
\includegraphics[width=2in]{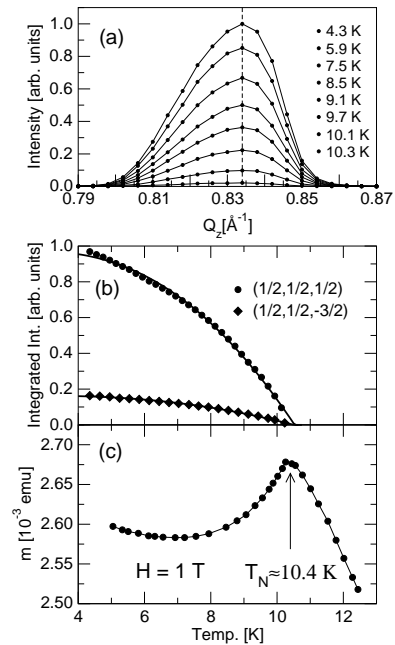}
\caption{\label{half-ord} $(a)$ Evolution of the
($\frac{1}{2}\frac{1}{2}\frac{1}{2}$) magnetic Bragg peak 
from 3~$\mu$m EuTe epilayer with the temperature. The asymmetry of the
peak indicates that the relaxation of the strain in the EuTe film is not 
complete.
$(b)$ Integrated intensity of the
($\frac{1}{2}\frac{1}{2}\frac{1}{2}$) and
($\frac{1}{2}\frac{1}{2}\frac{\overline{3}}{2}$) peaks vs. temperature
for the same epilayer.
$(c)$ The magnetic moment vs. temperature of 3~$\mu$m EuTe epilayer
measured by SQUID in applied magnetic field of 1 Tesla.}
\end{figure}

The dominant feature in diffraction spectra from Type II antiferromagnets is
a strong reflection at the ($\frac{1}{2}\frac {1}{2}\frac {1}{2}$)
position.
Pronounced maxima centered at the ($\frac{1}{2}\frac {1}{2}\frac {1}{2}$)
position were observed in all EuTe and EuTe/PbTe
SL samples cooled below the N\'eel temperature,
including  those in which the EuTe layer
thickness was as small as 2 monolayers.
The magnetic origin of the AFM SL peaks was confirmed in several
ways. First, the SL reflections also appear at other $Q$-space points
characteristic for the AFM II  structure, their intensities being
consistent with the Eu$^{2+}$ magnetic formfactor.
Secondly, the scattered intensity shows the typical temperature behavior,
closely following the squared Brillouin function for a $S=7/2$ system below
$T_{\text N}$. 
Finally, we have performed a number of diffraction measurements in
external magnetic fields that  convincingly prove the maxima, we have
investigated, originate from the ordering of the Eu magnetic moments and not 
from other effects that may potentially produce peaks at the
same reciprocal lattice points (such as, e.g., chemical ordering).

The maxima observed by us show that the Type II AFM ordering  was preserved 
in all EuTe layers and EuTe/PbTe SLs studied. Shown in 
Fig.~\ref{half-ord}$(a,b)$ is the temperature dependence of the magnetic 
peaks intensity for 3 $\mu$m EuTe epilayer sample. In Fig.~\ref{half-ord}$(c)$
the temperature dependence of the magnetic moment of another 3 $\mu$m EuTe
epilayer measured by a standard SQUID technique is presented \cite{holl-th}.
The critical temperatures  determined by neutron diffraction and magnetization 
measurements agree within the experimental errors. The obtained value of
$T_{\text N}=10.4 \pm 0.05$ K is slightly higher than 9.6 K found in bulk 
EuTe, due to the strain introduced to the epitaxial film by the BaF$_2$ 
substrate and PbTe buffer layer.
The distorted, non-gaussian profile of the $(\frac 12\frac 12\frac 12)$
magnetic Bragg peak in Fig.~\ref{half-ord}$a$ clearly points out to the
nonuniform lattice distortions present in the sample, the closer the portions 
of the EuTe film to the substrate the stronger the deformations of the
EuTe lattice. Thus in a sense, the 3 $\mu$m layer constitutes only a semi-bulk
 sample.
The influence of strain on magnetic properties of EuTe layers will be discussed
in detail in the following section.

\section{Strain and finite size effects}
\label{strain}

\subsection {$T$-domain structure}
\begin{figure}
\includegraphics[width=3.5in]{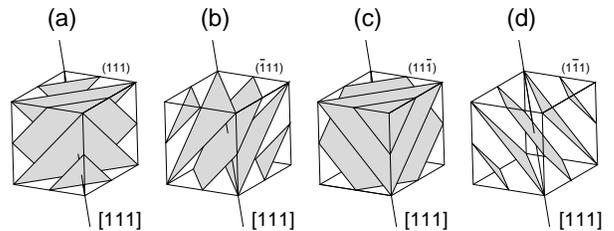}
\caption{\label{tdomain} The four symmetry equivalent Type II AFM arrangements
of the \{111\} ferromagnetic spin sheets in EuTe.}
\end{figure}

In a perfect FCC lattice there are four symmetry-equivalent Type
II AFM arrangements in which the FM spin sheets form on
the $(111)$, $(\overline{1}11)$, $(1\overline{1}1)$, or
$(11\overline{1})$ plane families (see Fig.~\ref{tdomain}).
These four configurations are usually referred to
as ``$T$-domains''. In a macroscopic,
strain-free EuTe crystal cooled through the N\'eel
point at $H_{\text{ext}}=0$ all four $T$-domain states would be
populated, giving rise to magnetic peaks at
$(\frac 12\frac 12\frac 12)$,
$(\overline{\frac 12}\frac 12\frac 12)$,
$(\frac 12\overline{\frac 12}\frac 12)$ and
$(\frac 12\frac 12\overline{\frac 12})$ reflection points. However,
in the EuTe epilayer and SL samples only the $(\frac 12\frac 12\frac 12)$ 
maximum
is observed in the neutron diffraction spectra; no detectable magnetic
scattering was ever found at the
other three reflection sites in any of the investigated specimens.
This means that in the layer systems only a single $T$-domain state forms
 -- the one in which the FM spin sheets are {\it parallel} to the
EuTe layers -- whereas the other three  `oblique' configurations never occur.

The  observed preference in the $T$-domain formation can be explained
by simple energy arguments. In the Type II AFM structure
a spin residing in a given FM sheet has six FM nearest neighbors (NNs)
located in the same sheet, and three AFM (`frustrated') NNs in each
adjacent sheet. The AFM next-nearest neighbors (NNNs) are also located
in the adjacent sheets, three in each.
Suppose that an (111) EuTe layer consists of $m$ monolayers.
Consider, e.g., an 'up' ($\uparrow$) spin located in the $i^{\text{th}}$
monolayer.
Table~\ref{tbl1} shows its neighbor environment  (i.e., how many of its
NNs and NNNs with parallel ($\uparrow$) and  antiparallel ($\downarrow$)
spins are located in the $(i-1)^{\text{th}}$, $(i)^{\text{th}}$,
and the $(i+1)^{\text{th}}$
monolayer)  for the `in-plane' and `oblique' domain configurations.
\begin{table}
\caption{\label{tbl1} Neighborhood spin configuration in an EuTe layer
for different $T$-domain types in which the ferromagnetic spin sheets are 
either parallel ('in-plane') or inclined ('oblique') to the (111) epitaxial
growth plane.}
\begin{tabular}{|c||c|c||c|c|}
\hline
Monolayer&\multicolumn{4}{|c|}{Domain type:}\\ \cline{2-5}
number:&
\multicolumn{2}{|c||}{`in-plane'}&\multicolumn{2}{|c|}{`oblique'}\\
\cline{2-5}
&NNs&NNNs&NNs&NNNs\\
\hline
$i-1$&3($\downarrow$)& 3($\downarrow$)&
2($\uparrow$)+ 1($\downarrow$)&
3($\downarrow$)\\
\hline
$i$&6($\uparrow$)& none
&2($\uparrow$)+ 4($\downarrow$)&none\\
\hline
$i+1$&3($\downarrow$)& 3($\downarrow$)&
2($\uparrow$)+ 1($\downarrow$)&
3($\downarrow$)\\
\hline
\end{tabular}
\end{table}
Because all NNNs in the Type II AFM structure are antiparallel,
the total energy of the NNN interactions is the same in the
`in-plane' and the `oblique' domains. The spins residing in
monolayers with numbers $2\le i\le m-1 $ have a full set of six
parallel and six antiparallel NNs, so that the total energy of the
NN interactions for these spins is zero, regardless of the domain
arrangement. However, the spins in the interface monolayers (i.e.,
those with $i=1$ and $i=m$) have only nine magnetic NNs:
6($\uparrow$)+3($\downarrow$) for the `in-plane' and
4($\uparrow$)+5($\downarrow$) for the `oblique' domains. Hence,
due to the finite thickness of the layer, the exchange energy of
the NN spins is not equal to zero any more and different for the
'in-plane' and 'oblique' configurations, with corresponding
average magnetic energy per Eu atom of :
\begin{equation}
\epsilon^{\text{in}} = 2C \left [ \frac{-3J_1}{m} + \frac{3(m-1)}{m}J_2
\right ]
\label{eq1}
\end{equation}
and
\begin{equation}
\epsilon^{\text{obl}} = 2C \left [ \frac{ J_1}{m} + \frac{3(m-1)}{m}J_2
\right ],
\label{eq2}
\end{equation}
where $C=S(S+1)/3$. Because $J_1>0$, the `in-plane' spin
arrangement is the one that minimizes the total magnetic energy.

Another factor that favors the `in-plane' domain configuration is
strain. Since $a_{\text{PbTe}}<a_{\text{EuTe}}$, the EuTe lattice
is `compressed' in the layer plane. Therefore the distance
$d_{\text{NN}}^{\text{$\parallel$}}$ between the NN Eu ions
located in the same (111) monolayer shortens,
while, due to the lattice reaction to the strain the distance
$d_{\text{NN}}^{\text{obl}}$ between the NNs residing in the adjacent
(111) monolayers increases. The
$d_{\text{NN}}^{\text{$\parallel$}}$ and
$d_{\text{NN}}^{\text{obl}}$ values may differ from the bulk NN
distance $d_{\text{NN}}^{\text{bulk}}$ by as much as $\pm 2$\%. As
shown by Goncharenko and Mirabeau \cite{goncha} from neutron
measurements under high hydrostatic pressure, the $J_1$ exchange
constants in Eu chalcogenides generally increases with the
decrease of the ion-ion separation. Hence, one can expect the
$J_1^{\text{$\parallel$}}$ for the `in-plane' NNs to be higher,
and the $J_1^{\text{$\perp$}}$ for the `out-of-plane' NNs
lower than $J_1^{\text{bulk}}$. However, for EuTe the rate of
$J_1$ change with $d_{\text{NN}}$ in the antiferromagnetic state
could not be measured in Ref.~\onlinecite{goncha} because the
contribution of $J_1$ to the N\'eel temperature cancels in the
cubic AFM state, i.e., $J_1$ is not accessible by experiment under
hydrostatic pressure. With respect to the NN exchange, $J_2$
appears to be only weakly dependent on the ion-ion separation, as
shown in Ref.~\onlinecite{goncha}.

Taking into account both the finite thickness and strain factors,
we obtain the average magnetic energy per Eu spin in EuTe layer
consisting of $m$ monolayers for the two different domain
arrangements in the form:
\begin{equation}
\epsilon^{\text{in}} = 2C \left [- 3\Delta J_1-
\frac{3J_1^{\text{$\perp$}}}{m} +
\frac{3(m-1)}{m}J_2 \right ]
\label{eq3}
\end{equation}
and
\begin{equation}
\epsilon^{\text{obl}} = 2C \left [ \Delta J_1+
\frac{J_1^{\text{$\perp$}}}{m} +
\frac{3(m-1)}{m}J_2 \right ]
\label{eq4}
\end{equation}
where $\Delta J_1 = J_1^{\text{$\parallel$}}-J_1^{\text{$\perp$}}$.

The first right-side term, which does not depend on the layer
thickness, reflects the effect of strain in the observed domain
type preference. The second term represents the finite thickness
effect as already given in Eq.~(\ref{eq1}) and (\ref{eq2}). In the thick layer 
limit ($m\rightarrow \infty$) the energy becomes 
\begin{displaymath}
\begin{array}{l}
\epsilon^{\text{in}}= 2C[-3\Delta J_1 + 3 J_2] \\
\epsilon^{\text{obl}}=2C[\Delta J_1 + 3 J_2]
\end{array}
\end{displaymath}
which for unstrained samples ($\Delta J_1 = 0$)
leads to the bulk value $\epsilon^{\text{bulk}} = 6CJ_2$ for both
arrangements. For small $m$ values, however, the difference
becomes significant (e.g., for $m=5$ and $m=3$ the energy per spin
for the in-plane domain arrangement is lower, respectively, by 8\%
and 35\% than for the oblique ones).

For the in-plane compressed EuTe layers  we introduce a parameter
which describes the ratio of the elongation of the `out-of-plane' NNs bonds
to the shortening of the distance between the `in-plane' NNs:
$k={\Delta d_{\text{NN}}^{\text{obl}}}/{\Delta d_
{\text{NN}}^{\text{$\parallel$}}}$. This parameter can be expressed in terms
of the in-plane $e_{\parallel}$ and out-of-plane $e_{\perp}$ strain, i.e.,
\begin{equation}
k=\frac{1}{e_{\parallel}}\left[\frac{1}{\sqrt{3}}
\sqrt{(1+e_{\parallel})^2 + 2(1+e_{\perp})^2}-1\right]
\label{eq5}
\end{equation}
For small strain values $e$, $k$ is essentially constant and
directly proportional to the $e_{\perp}/e_{\parallel}$ ratio.
With the value of the Poisson ratio for biaxially strained (111)
EuTe layers $\nu_{111}=0.301$ and the relation
$e_{\perp}/e_{\parallel}=2\nu/(\nu-1)$, we obtain  $k \approx -0.24$ 
for a compressive in-plane strain values of $e_{\parallel} \leq$ 2\%.

Using the $k$ parameter and denoting by $\xi$ the value
of $(\partial J_1/\partial d_{\text{NN}})$, we
can rewrite the magnetic energy
given by Eq.~(\ref{eq3}), in the form:
\begin{equation}
\epsilon^{\text{in}} = -6C \left [
\frac{J_1}{m} - \frac{(m-1)}{m}J_2 +
\frac{(m(k+1)-1)}{m}\xi d_{\text{NN}}\right ]
\label{eq6}
\end{equation}

In the  EuTe/PbTe structures the compressive strain and finite
thickness both appear to favor the `in-plane' $T$-domain
arrangement. It is interesting to note, however, that if the
spacer material had a larger lattice constant than EuTe and
produced a tensile strain in the EuTe layers, the strain would
favor the `oblique' arrangement and the two effects would compete
with each other. The domain arrangement in such superlattices
could be then tailored by manipulating the EuTe layer thickness.

\subsection {Changes of the N\'eel temperature}
\begin{figure}
\includegraphics[width=2in]{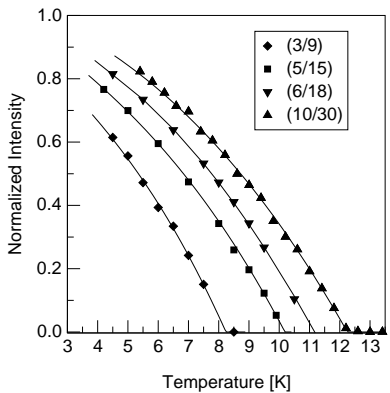}
\caption{\label{tneel-neut} The intensity of the $(\frac 12 \frac
12 \frac 12)$ magnetic diffraction peak for different EuTe/PbTe
superlattices. The solid lines correspond to the fit with the squared
mean field Brillouin magnetization function, with $T_{\text{N}}$ as
an adjustable parameter.}
\end{figure}

The effect of the strain and the finite thickness in the EuTe/PbTe
system is not only demonstrated by the preferred in-plane spin
alignment in the EuTe layers but also substantial shifts in the
N\'eel temperature. This is clearly indicated by the measurements
of the intensity of the $(\frac 12 \frac 12 \frac 12)$ magnetic
diffraction signal as a function of temperature as shown in
Fig.~\ref{tneel-neut} for several different SL samples with
different EuTe layer thicknesses $m$. The shapes of these curves
were found to be in good agreement with the squared mean-field
Brillouin magnetization function for $S= 7/2$ indicated by the
solid lines in Fig.~\ref{tneel-neut}. The transition temperature
$T_{\text{N}}$ was determined by fitting the function to the
measured data, with $T_{\text{N}}$ as an adjustable parameter. For
most samples the experimental $T_{\text{N}}$ values differs
significantly from the bulk value of $T_{\text{N}} = 9.6$ K (see
Fig.~\ref{tneel-neut}).

\begin{figure}
\includegraphics[width=2.in]{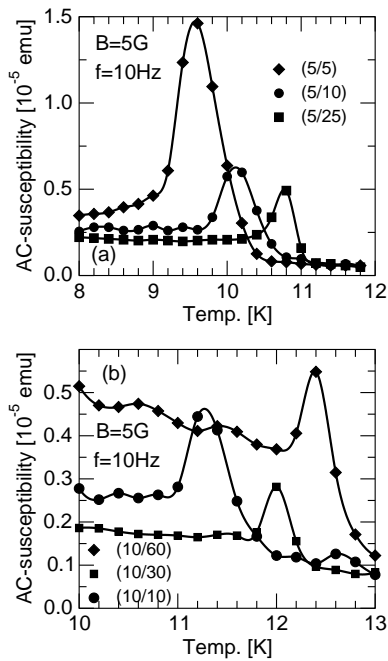}
\caption{\label{krenn-mag} Magnetic susceptibility of several EuTe/PbTe
superlattices measured with 10 Hz AC SQUID magnetometry.}
\end{figure}

The SL samples were also investigated by magnetometric
methods \cite{kepa02}. Examples of the magnetic susceptibility vs.
temperature dependence, which was measured using a 10 Hz AC SQUID 
magnetometer, are
shown in Fig.~\ref{krenn-mag} for samples with constant EuTe layer
thickness of $m=5$ (top panel) and $m=10$ (lower panel) but
varying PbTe spacer thickness $n$.
The observed $\chi (T)$ characteristics
for the SL specimens showed distinct maxima indicative for the
AFM phase transition. The N\'eel temperatures
obtained by this method were in good agreement (within
$\pm 0.4$ K) with the $T_{\text{N}}$ values yielded by neutron
diffraction experiments.

\begin{figure}
\includegraphics[width=2.5in]{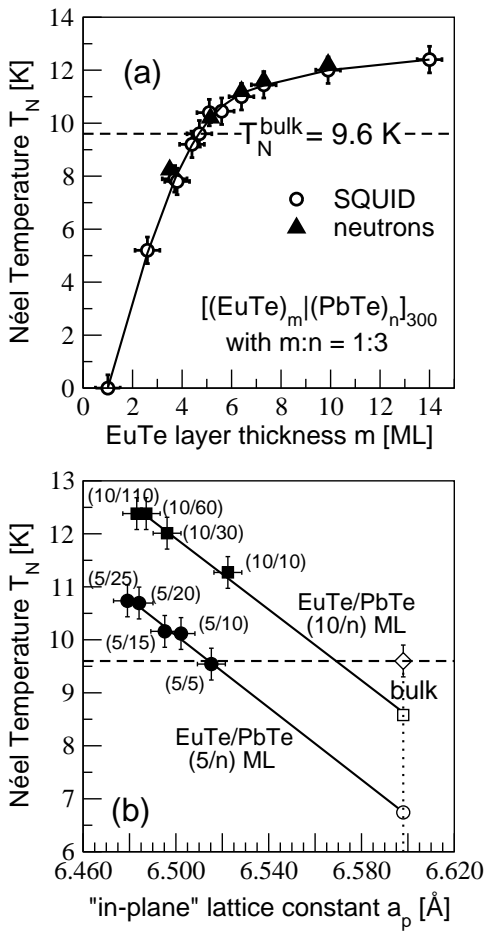}
\caption{\label{tn-strain} $(a)$ The N\'eel temperature for the
`constant strain' samples ($m$:$n$=1:3) vs. EuTe layer thickness. $(b)$ The
strain dependence of the N\'eel temperature for the two families
(5/$n$) and (10/$n$) of EuTe/PbTe superlattices. The solid lines are
only a guide for the eye.}
\end{figure}

The analysis of both, the neutron diffraction and magnetization data for a
large number of [(EuTe)$_m|$(PbTe)$_n$]$_N$
specimens with different $m$ and $n$ values reveals certain distinct trends in
the $T_{\text{N}}$ behavior. The experiments were carried out on series of
SL samples for which:
\begin{itemize}
\item[(i)] either the strain in the EuTe layers was approximately
constant (by keeping the ratio of $m$ to $n$ fixed), see Figs.~\ref{tneel-neut}
and \ref{tn-strain}$(a)$, or \item[(ii)] the EuTe layer thickness
$(m)$ was constant, and the PbTe spacer thickness  $(n)$ varied,
see Figs.~\ref{krenn-mag} and ~\ref{tn-strain}$(b)$.
\end{itemize}
In the `constant-strain' samples with very thin EuTe layers $(m=2)$
the $T_{\text{N}}$ was found to be considerably lower than
the bulk value 9.6 K,
but it increased with increasing $m$.
For $m \approx 5$ it exceeded the $T_{\text{N}}^{\text{bulk}}$,
showing a tendency to level out at a significantly higher value of 12.6 K
(Fig.~\ref{tn-strain}$(a)$).
For the structures with fixed EuTe layer thickness the $T_{\text{N}}$
showed a clear growing tendency when the PbTe spacer thickness was increased.
The larger value of $n$ for constant $m$ increases the strain in the
magnetic layers.  The in-plane strain $e_{\parallel}$
(or the in-plane lattice constant $a_{\parallel}$)
within the EuTe layers was determined by x-ray diffraction.
Plotted against $a_{\parallel}$, the
$T_{\text{N}}$ data from the sample series with constant $m$ show
approximately linear behavior (Fig.~\ref{tn-strain}$(b)$).

The trends observed in the $T_{\text{N}}$ behavior can be
explained on the grounds of the same simple model that has been used for
explaining the preference in the $T$-domain formation. Adopting the
mean-field theory approach, one can assume that the phase transition
temperature is proportional to the effective field experienced by an
`average' spin at $T=0$ -- in other words, to the average
energy per spin $\epsilon$ in the ground state,
given by Eq.~(\ref{eq3}). Taking into consideration
that $T_{\text{N}}/T_{\text{N}}^{\text{bulk}}=
\epsilon/\epsilon^{\text{bulk}}$, one obtains the expression for the
N\'eel temperature of thick strained layers:
\begin{equation}
T_{\text{N}}=T_{\text{N}}^{\text{bulk}}
\left [1+
\frac{\Delta J_1}{|J_2|} \right ]
\label{eq7}
\end{equation}
For small lattice distortions we assume that $\Delta J_1$ is
proportional to the distortion parameter and hence, $T_{\text{N}}$
should exhibit a linear dependence on the in-plane lattice
constant. From the linear fit to the $m=10$ data points in
Fig.~\ref{tn-strain}$(b)$ we obtain the rate of change of
$T_{\text N}$ with $a_{\parallel}$: $\partial T_{\text N}/\partial
a_{\parallel} = -33.8 $~K/\AA$ \pm 5\%$ and, correspondingly,
$\xi=0.41$~K/(\AA$ k_{\text B})$. For thinner layers, when the
$-6J_1^{\perp}/m$ term may not be neglected, this linear
characteristic should shift to the left. This is indeed consistent
with the $T_{\text N}$ vs. $a_{\parallel}$ behavior observed for
the $m=10$ and $m=5$ sample series.

To obtain an expression for  $T_{\text N}$ as a function
of the distortion parameter
$\Delta d_{\text NN}^{\parallel}$ and the layer thickness $m$, we
use as before the $T_{\text{N}}/T_{\text N}^{\text{bulk}}=
\epsilon/\epsilon^{\text{bulk}}$ relation with $\epsilon$ given by
Eq.~(\ref{eq6}):
\begin{equation}
\frac{T_{\text{N}}}{T_{\text{N}}^{\text{bulk}}}
=1 -\frac{1}{m} \left ( 1-\frac{J_1}{|J_2|}\right )
+\frac{\xi\Delta d_{\text{NN}}^{\parallel}((k+1)m-1)}{m|J_2|}.
\label{eq8}
\end{equation}
From Eq.~(\ref{eq8}) it follows that:
\begin{itemize}
\item[(i)] For samples with the same $m$, $T_{\text{N}}$
exhibits a linear dependence on
$\Delta d_{\text{NN}}^{\parallel}$.
\item[(ii)] The slope of the
$T_{\text N}$ vs. $\Delta d_{\text{NN}}^{\parallel}$ line
is equal to $\xi \Delta d_{\text{NN}}^{\parallel}(2m-1)/m|J_2|$;
it is weakly dependent on $m$, except for very thin layers, where
the change in the $(2m-1)/m$ factor becomes significant;
\item[(iii)]
With decreasing $m$,
$T_{\text{N}}$ shifts to lower values.
The `zero-strain' $T_{\text{N}}$
value is given by Eq.~(\ref{eq8}) by extrapolation to
$\Delta d_{\text{NN}}^{\parallel}=0)$.
\end{itemize}

The qualitative predictions of the above simple model appear to be consistent
with the $T_{\text{N}}$ vs. $\Delta d_{\text{NN}}^{\parallel}$
measured data for the $m=10$ and $m=5$ SLs series.
The rate of $J_1$ change yielded by the data
appears to be approximately 24\% per 1\% in the ion-ion distance change.
This result supports the conclusions by Goncharenko and Mirabeau \cite{goncha}
concerning the $J_1$ dependence on distance.  The calculated
`zero-strain' $T_{\text{N}}$ for $m=10$ and $m=5$ is 9.1 K and
8.6 K, respectively, whereas the corresponding values extrapolated from the
measured characteristics are 8.5 K and 7.0 K.
From Eq.~(\ref{eq8}) one obtains that the lowest possible N\'eel temperature is
7.2 K (for unstrained layers with $m=2$). Actually, experiments on
samples with $m=2$ yielded $T_{\text{N}}$ values close to 5 K.

This simple mean-field model correctly explains the qualitative
behavior of $T_{\text{N}}$ in layers with various thicknesses and
strain values. However, quantitatively the model appears to be
less successful, especially for very thin layers, showing 10\% to
20\% discrepancy between the model and experimental $T_{\text{N}}$
values. 
One possible reason of this discrepancy may be structural
imperfections that certainly exist in real superlattices. For
instance, as indicated by the results of magnetization
studies of EuTe/PbTe(100) SLs \cite{kostyk} and
of a similar SL system EuS/PbS \cite{stachow}, even slight
interdiffusion effects in the magnetic/nonmagnetic interface
regions may lead to observable decrease of the phase transition
temperature due to the reduction of the number of NN and NNN
exchange bonds between the Eu spins.

\subsection {$S$-domain structure and net magnetic moment}
\label{sdomain}
In the model outlined above it was assumed that the Eu spins in each 
individual EuTe layer form a perfectly homogeneous Type II AFM order.
The fact is, however,
that in the (111) layer plane there are three $\langle
11\overline{2}\rangle$ easy axes, 120$^\circ$ apart. This makes
possible six microscopically inequivalent domain arrangements (usually
referred as the $S$-domains \cite{oliv}).
It becomes a natural question whether each individual EuTe layer
in the SL structure constitutes a single $S$-domain, or
does it consist of many smaller ones, in which the spins are
oriented along different easy axes. Another important question
concerns the magnetic moment of the SL.
The layers can be thought of as truly
AFM only if $m$ is an {\em even} number.
For an odd $m$, however, the layer as a whole, or the
constituent $S$-domains should posses an uncompensated moment --
in other words, the layers become {\it ferrimagnetic}.
Note, that the opposite spin configurations in successive EuTe layers
could also lead to zero net magnetic moment of the entire SL, but this would 
require strong, perfect interlayer spin correlations.

Information about the $S$-domain structure and the net layer
moment is important for understanding the interlayer coupling
effects seen in the EuTe/PbTe SLs.  There is no direct
method of visualizing domains buried in a SL structure
or measuring their uncompensated moment.  Yet, much insight into
both these issues may be obtained from magnetization measurements
and from neutron diffraction studies of  spin rotation processes
in a magnetic field applied parallel to the EuTe layers.

\begin{figure}
\includegraphics[width=3in]{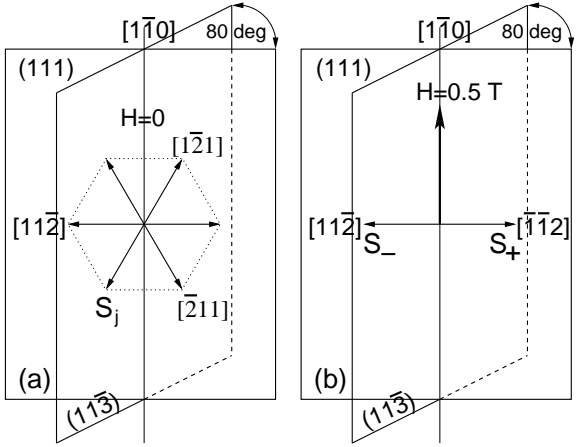}
\caption{\label{fldrot} Directions of Eu spins $S_j$ in zero external 
field $(a)$, and in the applied field $<1$ T $(b)$.}
\end{figure}

\begin{figure}
\includegraphics[width=3in]{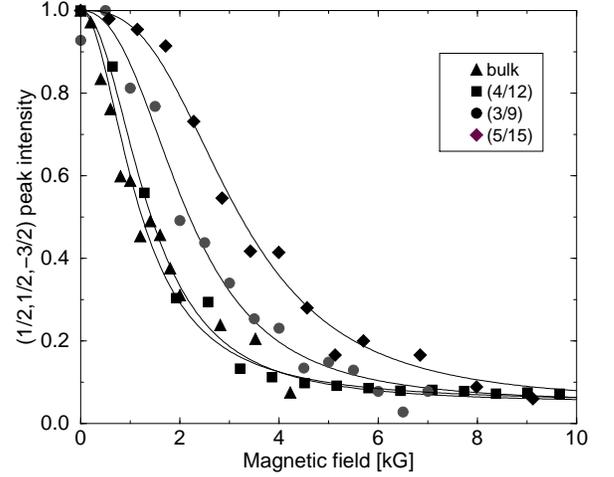}
\caption{\label{dmrotat} Intensity of the $(\frac{1}{2} \frac{1}{2}
\frac{\overline 3}{2})$  reflection vs. applied magnetic field for a bulk
EuTe specimen and several EuTe/PbTe SLs.}
\end{figure}

In standard measurements using an unpolarized neutron beam the magnetic
diffraction intensity is proportional to $\cos^2\alpha$, where $\alpha$
is the angle between the spins and the reflecting plane.
Since in the EuTe/PbTe
systems the spins lie in the (111) plane,
for the ($\frac 12\frac 12\frac 12$) reflection $\alpha=0$.
Any spin rotation in the (111) plane does
not change this value, and thus the reflection intensity does not change.
The information about spin orientation can be obtained from this reflection
only by using polarized incident neutrons and polarization analysis of the
diffracted beam. However, as the intensity in polarized neutron
experiments is typically about an order of magnitude or more lower than
in measurements with unpolarized beam, such studies appear to be too
time-consuming in the case of the EuTe/PbTe multilayers. Fortunately,
the same information can be obtained by studying the
($\frac 12 \frac 12\overline{\frac 32}$) magnetic reflection, taking
advantage of the fact that the reflecting plane associated with it,
$(11\overline{3})$, is nearly perpendicular to the (111) plane
($\arccos 1/\sqrt{33} = 80^\circ$), see Fig.~\ref{fldrot}.
In such experiments the external field $\vec{H}_{\text{ext}}$
is applied parallel to the
[$1 \overline{1} 0$] axis (i.e., the axis of intersection of the (111) and
($11\overline{3}$) planes), and the ($\frac 12 \frac 12\overline{\frac 32}$)
reflection intensity is measured vs. the field strength, as shown in
Fig.~\ref{dmrotat}.

First, let us discuss the anticipated outcome of such
an experiment based on the idealized picture of a EuTe layer.
Suppose that all possible $S$-domain states
are equally populated. Hence, when $H_{\text{ext}}=0$,
one-third of all spins are parallel to the
$[11\overline{2}]$ easy axis and make an angle $\alpha = 80^\circ$
with the
(11$\overline{3}$) plane. Two-third of the spins lie along the
$[\overline{2} 11]$ and the $[1 \overline{2} 1]$ easy axes
(Fig.~\ref{fldrot}$(a)$), and for them
$\alpha= 29.5^\circ$. Thus, the observed reflection intensity is
proportional to
$I \propto \frac 13 \cos^2 80^\circ +
\frac 23 \cos^2 29.5^\circ = 0.515$.

When the field is turned on, the system reaction should depend very much on
whether the number of spin monolayers is even or odd. Consider first an
even $m$, so that there is no uncompensated moment. In such a
situation, the field tends to turn the spins toward an orientation
perpendicular  to  $\vec{H}_{\text{ext}}$. It does
not affect the orientation of $\frac 13$ of the spins along the
$[11\overline{2}]$ axis which are already perpendicular to
$\vec{H}_{\text{ext}}$. To turn the remaining $\frac 23$ of the
spins, the field has to  overcome the anisotropy; the field strength
needed for that is
$H_{\text{ext}}\sim \sqrt{H_{\text a}H_{J}}$, where
$H_{\text{a}}$ is the anisotropy field, and $H_{J}$ is the
exchange field expressed in magnetic units. Since $H_{\text{a}}$
in EuTe has been found to be about 12 G at 4.2 K, and $H_{J}$
is about 3.5 T, the flop
to the perpendicular position should occur before $H_{\text{ext}}$
reaches 650 G. After that, all spins make an 80$^\circ$ angle with
the reflecting plane (see Fig.~\ref{fldrot}$(b)$), and the reflection
intensity is
$I \propto \cos^2 80^\circ = 0.03$; in other words, it should drop to
5.9\% of its zero-field value. With further $H_{\text{ext}}$
increase the intensity should not change much until
reaching the Tesla region, where the Eu moments start inclining
significantly toward the field, leading to further
suppression of AFM reflection. However, when the field is gradually
decreased, the original zero-field intensity should not be restored,
because all the  spins should remain `locked' in the perpendicular
position by the anisotropy field.

This idealized model scenario changes quite dramatically for an
odd $m$. Now, the field tends to align the uncompensated moment
parallel to  $\vec{H}_{\text{ext}}$. The field strength needed
to overcome the anisotropy is
$\sim m H_{\text{a}}$, i.e., $\leq 100$ G
for the $m\leq 10$ systems studied by us. In such
$H_{\text{ext}}$, all the spins get aligned
parallel to the (11$\overline{3}$) plane, so now $\alpha = 0$ and
$I\propto \cos 0 = 1$; in other words,
in $H_{\text{ext}} \approx 100$ G,
there should be an increase of the
($\frac 12 \frac 12\overline{\frac 32}$) reflection intensity by a factor
of almost two. After that initial jump, the intensity should not change
until the field reaches a value where inclining of the Eu moments
from the magnetic field lowers the system energy.
After the field returns to zero, the spins should chose easy axes
{\em nearest} to the field direction, so that the $S$-domains
corresponding to the $[11\overline{2}]$ direction should not be
repopulated. All spins now make a 29.5$^\circ$ angle with the
reflecting plane, so that $I \propto \cos 29.5^\circ = 0.8$
should be about 60\% higher than originally.

As illustrated in Fig.~\ref{dmrotat}, the significant increase of the
reflection intensity, expected to occur in samples with odd $m$, was never
observed in any real EuTe/PbTe SL.
A slight increase ($\sim 10$\%)  was observed
in some specimens  -- however, for samples with even $m$.
Moreover, the observed $I(H_{\text{ext}})$ characteristics are almost
completely reversible -- there is no indication for any
hysteresis. After the field was  increased and then returend
back to zero, in all samples the original reflection intensity
was restored within  experimental error.

All studied samples behaved essentially as in the `even-$m$ scenario'
(apart from reversibility), thus showing no uncompensated magnetic moment.
However, the field needed to rotate all spins to the direction
perpendicular to ${H}_{\text{ext}}$  was always significantly higher than 
the expected value of
$H_{\text{ext}}=\sqrt{H_{\text{a}}H_{J}} \approx 650$ G, corresponding to the
bulk value of $H_{\text a}=12$ G. Moreover, the
shape of the curves in Fig.~\ref{dmrotat} is not consistent with a 
single $H_{\text a}$ value, but rather suggest that in each sample there is
a statistical distribution of the anisotropy fields.
In fact, a satisfactory description of the observed
curves was obtained by assuming a Gaussian
distribution of the $H_{\text a}$ values. The mean $H_{\text a}$ values
obtained from the fits for different samples
varied from 50 to 200 G, but no systematic trend has been found.
To comment on the fact that no traces of any hysteresis were observed in the 
studied samples, we recall that the expected `even-$m$ scenario' 
irreversibility was deduced for idealized, perfect
layers with the three-fold symmetry of anisotropy fields as 
in the bulk material.
In real SLs, not only are the values of $H_{\text a}$  different,
as obtained from the fits in Fig.~\ref{dmrotat}, but also distribution of
their directions may deviate substantially from the <211> axes due to the
influence of various types of defects and inhomogeneities.

The absence of the ferrimagnetic properties in the SL samples with nominally
odd number $m$ of magnetic monolayers, seen in
the ($\frac 12 \frac 12\overline{\frac 32}$)
reflection intensity vs. field strength measurements, was further
corroborated by the magnetization studies.
\begin{figure}
\includegraphics[width=3in]{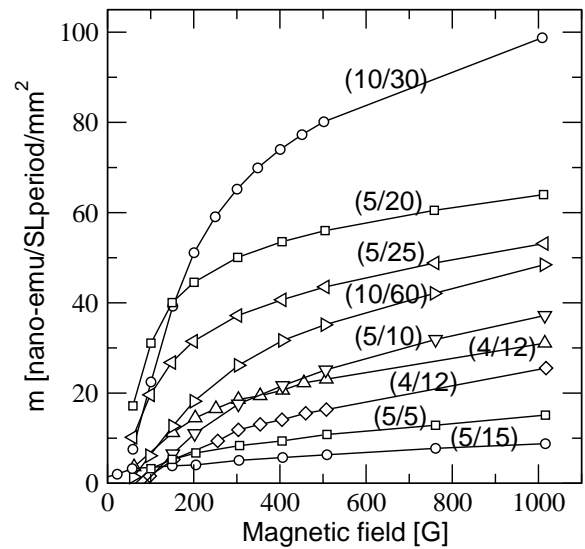}
\caption{\label{stefan} Magnetic moment per SL period, per mm$^2$ for several
EuTe/PbTe SLs with nominally odd and even number of magnetic monolayers in
EuTe layer vs. applied magnetic field.}
\end{figure}
In Fig.~\ref{stefan} the magnetic moments for various samples,
per SL period, per mm$^2$, are plotted vs. the magnetic field up to 1 kG,
which should align all the uncompensated spins parallel to the field.
As shown in Fig.~\ref{stefan}, for all SL samples, with both nominally even 
and odd $m$, the magnetic moment is several
times smaller than the value 340 nanoemu one should observe
from the 1 mm $^2$ of an uncompensated EuTe monolayer.

\begin{figure}
\includegraphics[width=3in]{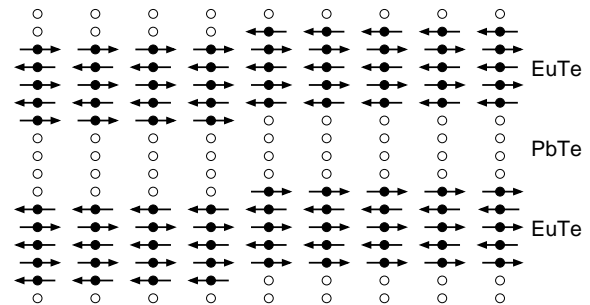}
\caption{\label{terr} Diagram of (5/4) EuTe/PbTe SL with conformally 
repeated terrace structure, that leads to a large reduction  of the total
magnetic moment of the individual odd-$m$ EuTe layer as compared with 
an identical layer without such a step. For clarity, only the cations, Eu
(solid circles) and Pb (open circles), are presented and the anion Te
atoms are omitted.}
\end{figure}

The observed field dependence of the intensity of the ($\frac 12 \frac
12\overline{\frac 32}$) peak, the absence of hysteresis and the low
values of the net magnetic moments at intermediate magnetic fields all
indicate that the vanishing of the magnetic moments in the SLs
with nominally odd numbers of Eu monolayers can neither result from a
random orientations of ferrimagnetic domains, nor from an antiparallel
orientation of such domains in consecutive EuTe layers due to interlayer 
coupling. It must result
from almost complete compensation of the magnetic moment within
each domain. Such an unexpected compensation may be attributed to,
e.g., a specific terrace structure of EuTe layers with one
monolayer steps as shown in Fig.~\ref{terr}.
The existence of one monolayer thick steps on the surface of MBE deposited 
EuTe layers has been confirmed by scanning tunneling microscopy 
studies \cite{springh2}.
The X-ray and neutron diffraction
spectra, reported in the next section, prove the very high
structural quality of our SLs, without traces of any significant
interface roughness. They do not exclude, however, such terrace
structures, the more so when the steps are conformally repeated
over several SL periods.

\section{Interlayer coupling}
\label{coupling}
\subsection {Neutron diffraction in zero magnetic field}
\begin{figure}
\includegraphics[width=3.5in]{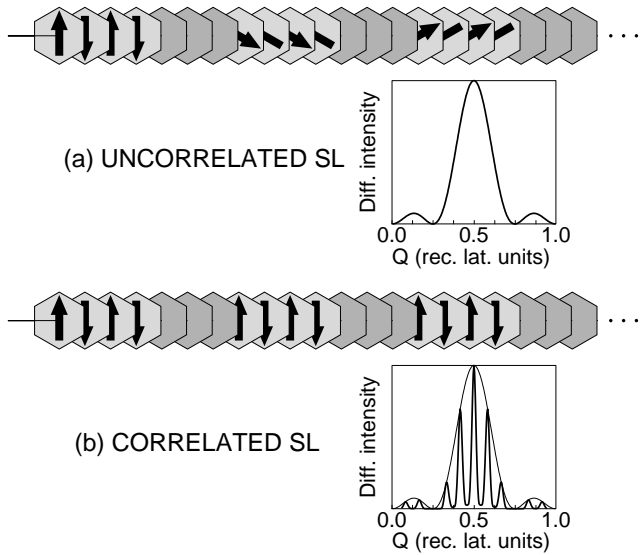}
\caption{\label{cor} Possible spin configurations in EuTe/PbTe SLs.
$(a)$ Uncorrelated SL; directions of the monolayer
magnetizations in consecutive EuTe layers change randomly giving rise to a
single broad maximum in a neutron diffraction pattern. $(b)$ Correlated SL;
directions of the monolayer magnetizations in consecutive EuTe layers change
in a regular way; in the case presented in the Figure, the
orientations of the spins in all layers are the same. Corresponding
diffraction pattern exhibits a number of narrow fringes.}
\end{figure}

\begin{figure}
\includegraphics[width=3in]{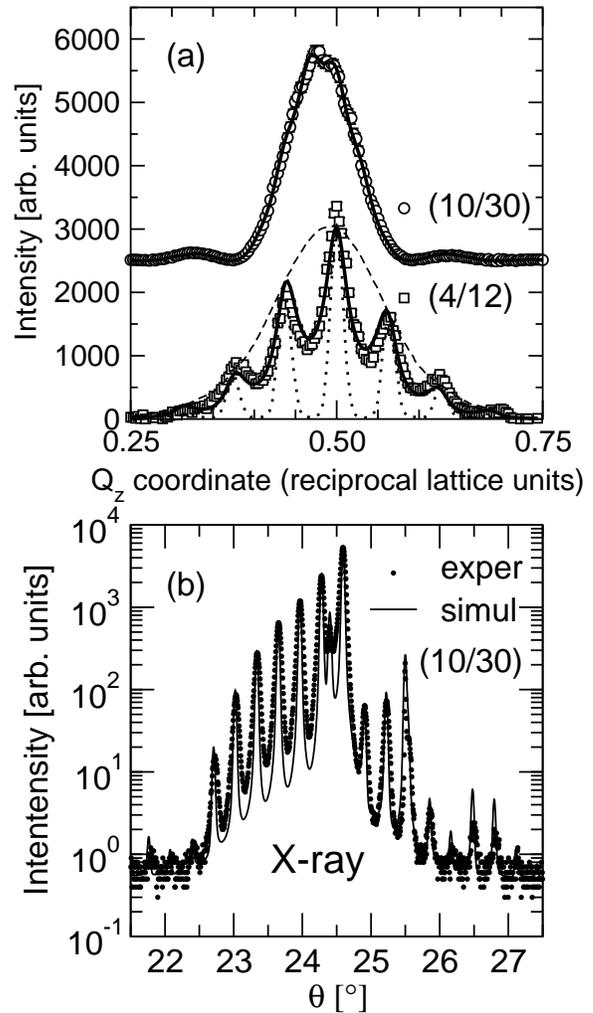}
\caption{\label{1030nx}$(a)$
Neutron diffraction scans along the [111]
direction through the  principal ($\frac12 \frac 12 \frac 12$)
AFM reflection in a 10/30 and 4/12 SL sample, showing
pronounced satellite peaks for the latter one. The dotted line represents
the expected shape of the spectrum for the fully correlated SL, with the 
instrumental resolution taken into account. The dashed line shows the single
layer magnetic structure factor.  $(b)$
High-resolution X-ray diffraction scan through the (222)
reflection in the 10/30 SL specimen (points: measured data, line:
dynamical simulation).}
\end{figure}

\begin{figure}
\includegraphics[width=3in]{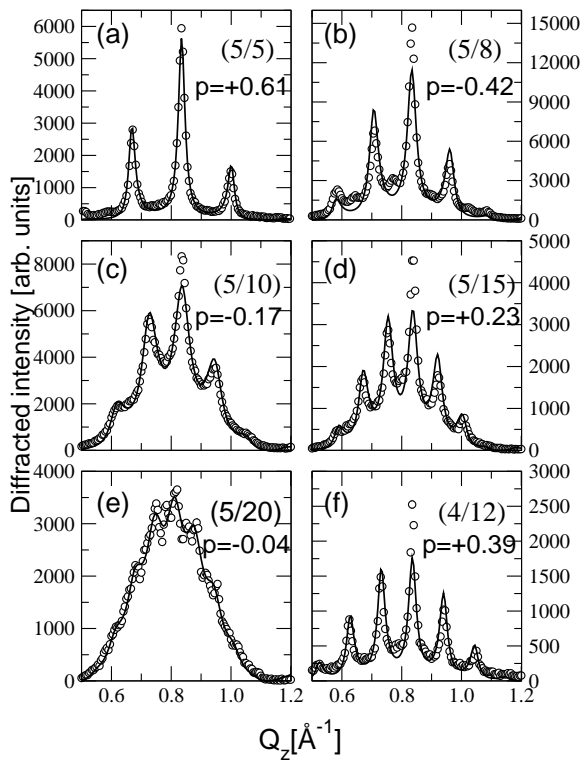}
\caption{\label{correl0} High resolution magnetic diffraction patterns from
several SL samples measured with NG-1 reflectometer. 
The solid curves are fits of Eq.~(\ref{partial}) to the data points. 
The fitted values of the partial correlation coefficient $p$ for each spectrum 
(as defined in subsection D) is shown in the figure. The small additional 
peaks visible in between the magnetic SL satellites must not be
attributed to the correlations of opposite sign - they can as well result
from different periodicity in a small portion of the SL (compare Appendix).}
\end{figure}

Magnetic neutron diffraction is the only experimental tool capable of
revealing the interlayer spin correlations in the case of AFM/nonmagnetic
multilayers. The principle of the method is illustrated in Fig.~\ref{cor}.
Neutron diffraction scan along the [111] direction through the
($\frac 12 \frac 12 \frac 12$) reflection point
from [(EuTe)$_{10}|$(PbTe)$_{30}$]$_{300}$
specimen with rather large thickness of the nonmagnetic PbTe
spacers ($d_{\text{PbTe}} =$ 112\AA) is displayed in Fig.~\ref{1030nx}$(a)$.
The large number of SL satellite peaks in the X-ray diffraction
spectrum and the excellent agreement with dynamical simulations
(Fig.~\ref{1030nx}$(b)$) prove that the structural quality of the
specimen is very high. In contrast to the multipeaked X-ray
pattern, the magnetic neutron diffraction spectrum of this
SL has only the form of a single broad peak accompanied
by two weak subsidiary side maxima. This profile shows a close
similarity to the squared structure factor of a {\it single}
(EuTe)$_m$ layer, $|F_{\text{BL}}|^2$, known from the standard
diffraction theory \cite{erwin} (also see the Appendix). 
Such a spectrum shape produced by a {\it
multilayer} structure indicates the lack of coherence between the
waves scattered by successive layers, meaning that the spin
alignments in these layers are not correlated. However, when the
PbTe spacer thickness decreases, the character of the AFM
reflections dramatically changes. As exemplified in Fig.~\ref{1030nx} for
the 4/12 SL, a distinct pattern of narrower
satellite peaks then emerges at regular intervals $\Delta Q_z$
equal to the spacing between the satellite peaks in the X-ray
spectra. This clearly indicates the formation of magnetic
interlayer correlations across the PbTe spacers. For
$d_{\text{PbTe}}$ below $\sim 60$ \AA{} these magnetic satellites
become the dominant part of the spectrum, as is also shown in
Fig.~\ref{correl0} for a series of [(EuTe)$_5|$(PbTe)$_{\text
n}$]$_{300}$ SL samples with varying PbTe spacer thickness.

\subsection {Neutron diffraction in high magnetic field}
\begin{figure*}
\includegraphics[width=6in]{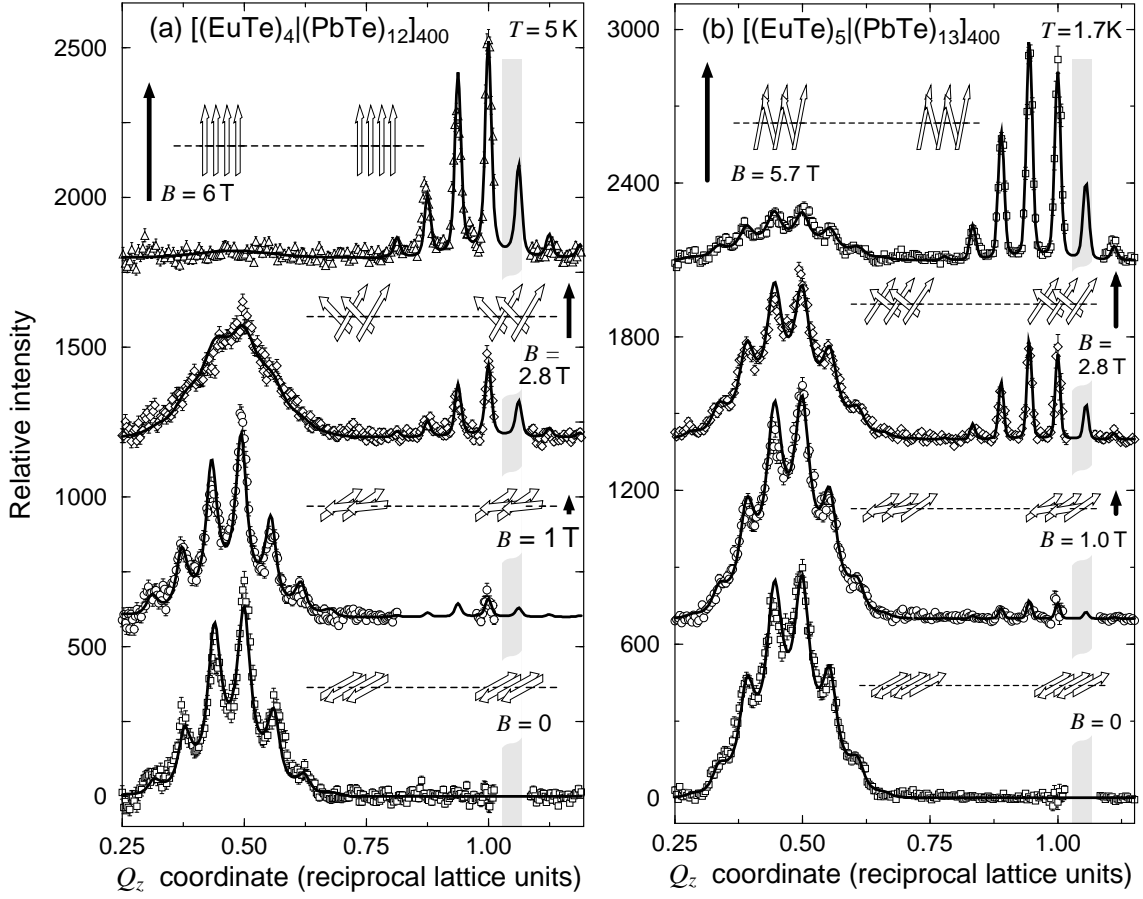}
\caption{\label{hifield} Extended neutron diffraction scans
for (4/12) $(a)$ and (5/13) $(b)$  SL samples,
showing the suppression of the AFM scattering at the 
$(\frac 12 \frac 12 \frac12)$ reflection point $(Q_z=0.5)$ with increasing
external magnetic fields, and the emerging of new satellite lines near
the (111) reciprocal lattice point $(Q_z=1.0)$  due to the induced FM 
spin alignment
(chemical structure contributions are subtracted -- the data
represent purely magnetic scattering). The shaded bars (removed data points)
indicate the regions where a very strong reflection from the BaF$_2$
substrate occurs.
The solid curves are the best fits of Eq.~(\ref{partial}) to the data.
The corresponding alignments of magnetization in successive spin monolayers
are shown by the arrows.}
\end{figure*}

The evolution of the ($\frac{1}{2}\frac{1}{2}\frac{1}{2}$)
reflection for two, [(EuTe)$_4|$(PbTe)$_{12}$]$_{400}$ and
[(EuTe)$_5|$(PbTe)$_{15}$]$_{400}$, SLs in  $\vec{H}_{\text{ext}}$
parallel to the [1$\overline1$0] axis is presented in
Fig.~\ref{hifield}. With increasing $H_{\text{ext}}$,
antiferromagnetically coupled spins first rotate towards
directions perpendicular to the field and then gradually incline
towards the field (see schemes in Fig.~\ref{hifield}).
Accordingly, the AFM ($\frac 12\frac 12\frac 12$) diffraction
structure gradually fades away, while a new set of peaks emerges
in the (111)  FM diffraction region at $Q_x = 1$. The almost total
disappearance of the AFM component at 6 T is consistent with the
behavior of bulk EuTe in external magnetic fields \cite{oliv}.
With respect to the magnetic interlayer coupling, it is crucial to
note, that the FM SL peaks at $Q_z = 1$ are quite narrow (only
slightly broadened beyond the instrumental linewidth) as compared
to the significantly broader satellite peaks in the AFM region at
$Q_z = 1/2$ and $B_{\text{ext}}=0$. In addition, there is no broad
background in the FM region. The sharp satellite peaks in the FM
state are due to the perfect long range spin coherence due to spin
alignment by the high external magnetic field, which gives
additional clear evidence for the excellent structural quality of
the samples. As a consequence, the broadening  of the satellite
peaks in the AFM region {\it cannot} be attributed to structural
imperfections, but must be due to a limited long range {\it spin
coherency} between the magnetic layers in the AFM state that may
be induced only by spontaneous magnetic interlayer interactions.

\subsection {Field cooled samples}
\begin{figure}
\includegraphics[width=2.5in]{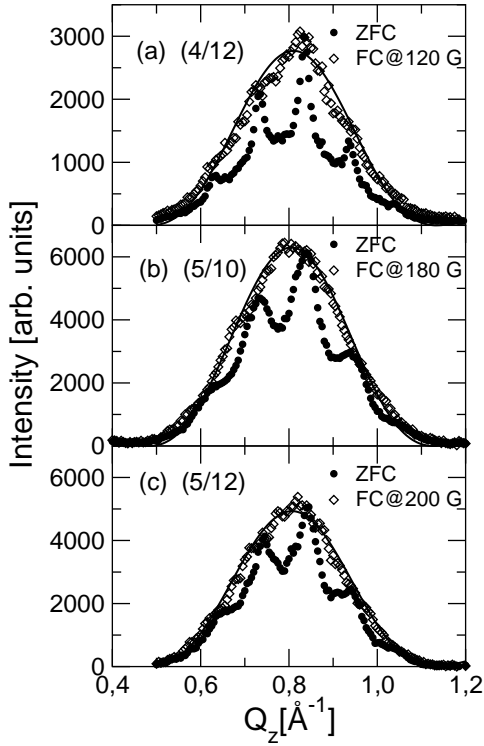}
\caption{\label{fcool} Neutron diffraction scans performed at
($\frac12 \frac 12 \frac 12$) reciprocal lattice position for the
zero-field-cooled (ZFC) and field-cooled (FC) (5/10), (5/12), and (4/12)
SLs. The FC data has been rescaled to show the effect of
enveloping the ZFC data. These diffraction data were collected on
the NG-1 reflectometer. The absence of the interlayer correlations
in the samples after field cooling is evident (some traces
are still visible in the 4/12 sample). For comparison the
calculated single layer structure factor is plotted with a
continuous line.}
\end{figure}

All the above experiments have been carried out on samples which
were cooled to temperatures below $T_{\text N}$ in zero external
magnetic field. In this case, the magnetic field applied to a
sample that is already in a correlated state does not destroy the
existing interlayer correlation unless the field becomes  strong
enough to influence the AFM order within the
individual layers. This takes place only for external fields much
stronger than 1 T. In contrast, cooling the samples from above to
below $T_{\text N}$ in relatively weak magnetic fields (of the
order of 100 - 200 G) almost entirely prevents the formation of
any interlayer correlations. This is demonstrated in
Fig.~\ref{fcool}, where the magnetic diffraction patterns obtained
for zero-field-cooled (ZFC) and the field-cooled (FC)
superlattices are depicted. The magnetic field was applied
parallel to the SL growth plane and oriented 
along the $[1 \overline{1} 0]$ crystallographic axis.
In the field-cooled cases (open symbols), the spectra only have
the form of the structure factor of a single layer
$|F_{\text{BL}}|^2$, characteristic for uncorrelated SLs (see
Fig.~\ref{cor}$(a)$), whereas the spectra obtained after zero-field cooling
show the usual satellite peaks attributed to interlayer
coupling (as shown in Fig.~\ref{cor}$(b)$). This change of 
the diffraction spectra just by the
application of an external field during cooling again demonstrates
the purely magnetic origin of the multi-peak structure.
Detailed experiments, in which the ($\frac12 \frac 12 \frac 12$) peak
profile was studied after cooling the sample in
different external fields show that the loss of correlations
exhibits a gradual dependence on the field strength.
The multi-peak spectrum
starts to evolve towards a broad maximum already for cooling at
fields as low as 10 G, but the final uncorrelated state is
being reached only when cooling at higher fields, usually a few
hundred gauss. However, the effect is fully reversible --
subsequent warming up and cooling down the samples again in zero
field restores the original correlated state. All the samples
under investigation have shown this type of behavior.

The possible explanation of the different behavior of the FC and
ZFC samples will be given in the next subsection. Here, however,
it should be emphasized that such a behavior excludes the possibility
that the satellite structure of the
($\frac12 \frac 12 \frac 12$) peak results from the
pinholes in the PbTe spacer and the formation of EuTe bridges across
the spacer region, because in such case the resulting interlayer coupling
would not depend so sensitively on the applied external fields.

\subsection{Comparison of the experimentally determined interlayer 
correlations with theoretical predictions.}
\label{theory1}
Three mechanisms of the interlayer correlations, which can be relevant 
to AFM/nonmagnetic semiconductor layer structures, have been proposed in 
literature. First, the transfer of the magnetic order to the next magnetic 
layer via the spin-polarized carriers bound to the impurities located in the 
spacer, was considered in Refs.~\onlinecite{{shevch, rusin}}. This does not 
seem to apply to EuTe/PbTe SLs, since in PbTe the large dielectric constant 
and small carrier effective masses prevent the formation of shallow impurity 
centers\cite{heinrich}. Another mechanism considered is the long-range dipolar 
interaction, which was investigated in Ref.~\onlinecite{borchers} for FM 
metallic layer systems with domain structure. It does not exist for perfect 
AFM layers, but one can argue that in real SLs the dipolar coupling between 
local magnetic moments related to interface terraces and steps as invoked 
in Section~\ref{sdomain}, can be 
effective. We expect, however, this mechanism to be much weaker here than in 
the FM case, since the dipole-dipole interaction is proportional to the square 
of the average dipole moment. In addition, we expect that such mechanism 
should be much more effective for SLs with odd number of magnetic monolayers 
than for those with $m$ even, where the terraces do not lead to local dipole 
moments. No such preference was observed in the experimental data. 

\begin{figure}
\includegraphics[width=3in]{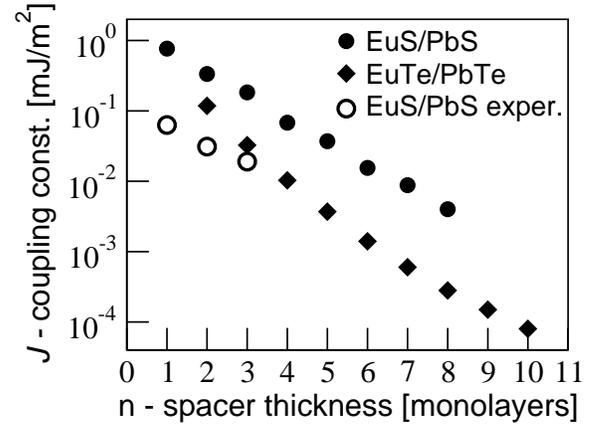}
\caption{\label{theory} The interlayer exchange constant $J$, defined by 
the  difference of the total electronic energy for 
the same and opposite spin configurations in successive magnetic layers of 
the SL, as a function of the spacer thickness for FM EuS/PbS and AFM
EuTe/PbTe SLs. Open circles represent the experimentally obtained $J$ 
for FM EuS/PbS. For the AFM SLs the direct experimental
determination of the strength of the coupling is not possible.}
\end{figure}

Finally, in Ref.~\onlinecite{blin} a mechanism was presented, which attributes 
the interlayer coupling to the sensitivity of the SL electronic energies to 
the magnetic order in consecutive magnetic layers.  The total energy of the 
valence electrons for two different magnetic SLs, one with the same and the 
other with opposite spin configurations in neighboring magnetic layers, was 
compared. The difference between these two energies was considered as a 
measure of the strength of the interlayer magnetic coupling resulting from 
band structure effects. In Ref.~\onlinecite{blin}
it was shown that in both studied types of IV-VI semiconductor structures, 
i.e., in FM EuS/PbS and AFM EuTe/PbTe, this mechanism can be effective.
As shown in Fig.~\ref{theory},
the calculated strength of the coupling decreases monotonically with the 
increasing thickness $n$ of the spacer and is 
practically independent of the thickness $m$ of magnetic layers.
In addition, it was 
found that for all FM and AFM SLs, regardless of their $m$ and $n$ values, the 
lower energy corresponds to the antiparallel alignment of the two spins facing 
each other across the spacer layer. In other words, the mechanism leads to an  
antiferromagnetic coupling between the FM layers. In the AFM EuTe/PbTe 
structures, however, the energetically preferred spin configuration along the 
SL growth axis depends on the parity of the number of monolayers within the
magnetic layer, i.e., the actual magnetic period is equal to the 
chemical period for even $m$ but twice as large for odd $m$.  

These theoretical results have proven to explain the experimental 
observations in the 
FM (001) EuS/PbS SLs \cite{kepa_epl}. For FM structures the 
magnetization and neutron reflectivity measurements in external magnetic 
fields, enable one to determine directly the strength of the coupling 
and compare it with the model. The sign of the 
interlayer exchange coupling and the rate of its decrease with the PbS 
nonmagnetic spacer thickness are in very good agreement with the predictions 
of the model, as shown in Fig.~\ref{theory}. 
The fact that the experimental values of the exchange constants estimated 
from the saturation fields in real FM structures are about an order of 
magnitude 
smaller than the theoretical ones was attributed in Ref.~\onlinecite{kepa_epl}
to the interfacial roughness and interdiffusion, which were shown to reduce 
significantly the strength of the interlayer coupling, also in metallic 
structures. 

For the AFM EuTe/PbTe structures the comparison of the theoretical 
predictions with the experimental data is much more complicated than 
for the FM structures -- in this case not only the perfect tool to measure the 
strength of the interlayer coupling, i.e., the saturation magnetization, is 
not applicable, but, as shown below, the correlated spin configurations are 
much more sensitive to the morphology of the SL. 

In our superlattices evidence for the interlayer coupling between the
AFM EuTe layers comes from the satellite structure of the
neutron diffraction spectra. The sign of the coupling can be determined only 
by a detailed analysis of the positions of the satellite peaks.  Moreover, to 
describe the observed shapes of the AFM diffraction spectra, we have to invoke 
the idea of ``partial correlations'', described by an interlayer correlation 
parameter $p$ $(|p|<1)$, as presented in detail in the Appendix.
An idealized fully correlated EuTe/PbTe SL would 
contain a single $S$-domain, with the monolayer magnetization sequence in 
any $i^{\text{th}}$ layer either {\em repeated} in the $(i+1)^{\text{th}}$ 
layer (perfect correlations with $p=+1$)
\begin{itemize}
\item []
$\uparrow\downarrow\uparrow
\downarrow\uparrow\cdots\cdots\uparrow\downarrow\uparrow
\downarrow\uparrow\cdots\cdots\uparrow\downarrow\uparrow
\downarrow\uparrow\cdots\cdots\uparrow\downarrow\uparrow
\downarrow\uparrow,$
\end{itemize}
or {\em reversed} in the $(i+1)^{\text{th}}$ layer (perfect correlations 
with $p=-1$)
\begin{itemize}
\item []
$\uparrow\downarrow\uparrow\downarrow\uparrow\cdots\cdots
\downarrow\uparrow\downarrow\uparrow\downarrow\cdots\cdots
\uparrow\downarrow\uparrow\downarrow\uparrow\cdots\cdots
\downarrow\uparrow\downarrow\uparrow\downarrow.$
\end{itemize}
In both cases, the AFM neutron diffraction pattern 
should exhibit a series of
very narrow peaks, with the width defined by the instrumental resolution only, 
as seen for the satellite peaks in the FM region at 
$H_{\text{ext}}$ = 6~T in Fig.~\ref{hifield} and Fig.~\ref{cor}(b). 
However, in real samples, this perfect long range SL order may be disrupted 
by ``phase lapses'' (i.e., switches to the other
S-domains types)  occurring at random intervals.
Such ``partially correlated'' 
chains can then be characterized by {\em fractional} $p$ values, now 
expressing the probability $P= (1+p)/2$ that any two adjacent EuTe layers 
have identical spin sequences.
Applying diffraction theory to such a system, one obtains 
the following expression for the magnetic diffraction intensity
(see the Appendix for the formula derivation):
\begin{equation}
I(Q_z) \propto |F_{\text{BL}}(Q_z)|^2 \frac{1-p^2}{1 -2p\cos(Q_{z}D)+p^2}
\label{partial}
\end{equation}
where $D=m\delta_{\text{EuTe}} + n\delta_{\text{PbTe}}$ is the SL 
period; $\delta_{\text{EuTe}}$ and $\delta_{\text{PbTe}}$ being the monolayer 
thicknesses of EuTe and PbTe,
respectively. This expression is similar to that used for analyzing
x-ray diffraction patterns from partially ordered layered 
structures \cite{hentell}.
\begin{figure}
\includegraphics[width=3in]{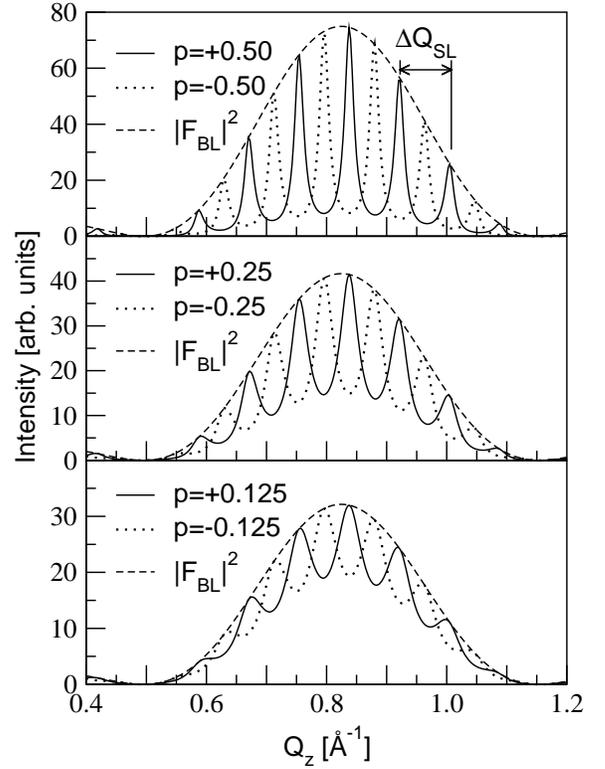}
\caption{\label{telhen} Influence of the magnitude and the sign of the 
correlation parameter $p$ on the magnetic diffraction intensity from
5/15 SL, as calculated from Eq.~(\ref{partial}). The change of the sign of $p$ 
leads to the shift of the
SL satellite peaks by half of the spectrum periodicity. With the decreasing
value of $|p|$ the width of the peaks increases and so does the height 
of the underlying ``hump''. The single EuTe layer magnetic structure factor
$|F_{\text{BL}}|^2$, enveloping the whole spectrum, is presented with the
dashed line.}
\end{figure}
The value of $p$ determines both the widths of the AFM satellite peaks  
as well as the height of the underlying ``hump" (see Fig.~\ref{telhen}). 
By adjusting the $p$ parameter for each sample
the observed spectral shapes are reproduced remarkably well (solid lines 
in Fig.~\ref{correl0}). The least-square fitted $|p|$ values are considerably 
lower than unity, even for relatively thin PbTe spacers. 
This indicates the presence of 
mechanisms inhibiting the correlation formation in the AFM EuTe/PbTe SLs. 
Although one can speculate that several effects, such as thickness 
fluctuations or interface roughness, may participate in the suppression, 
neutron diffraction measurements in moderate external in-plane 
fields $H_{\text{ext}} < 1$ T imply that magnetic 
anisotropy fields $H_{\text a}$ 
in random directions parallel to the layers play a major role in this effect. 
Two $S$-domains in EuTe layers facing each other across the nonmagnetic 
spacer may  become correlated only if the interlayer coupling energy is 
sufficiently high to overcome the anisotropy in at least one of them.

This observation together with the model described above
offers a simple qualitative explanation of the behavior of the field cooled 
samples shown in Fig.~\ref{fcool}. It is based on the 
fact that the interlayer coupling becomes effective only below the N\'eel 
temperature, when the AFM order in EuTe layers is already well 
established and the anisotropy fields are still weak. The interlayer coupling 
energy resulting from band structure effects is proportional to the cosine 
of the angle $\theta$ between the spins at the opposite borders of the 
nonmagnetic spacer. Thus, the torque responsible for the relative rotations 
of the spins in neighboring  EuTe layers should be proportional to 
$\sin\theta$. During cooling of the sample in an external magnetic field, 
all the 
spins align along the direction perpendicular to the magnetic field.
Thus, the torque is equal to zero and the interlayer correlations cannot be 
formed. In contrast, in the ZFC samples the AFM ordered spins in different 
magnetic layers align randomly along different 
in-plane directions, hence $\sin\theta$ and the torques are in 
general not equal to zero. Therefore, the rotation mechanism can be effective, 
leading to the correlated layer structure in the case of ZFC samples.

From Eq.~(\ref{partial}) and Fig.~\ref{telhen} it follows that for a 
given combination of 
$m$ and $n$ values the change of the $p$ sign corresponds to a half 
period shift $\frac 12\Delta Q_{\text SL}$ in the AFM satellite positions.  
For $Q_z$'s in the the vicinity of $(\frac12 \frac 12 \frac 12)$ reciprocal 
lattice point the same is 
expected when the $p$ is fixed, but either $m$ or $n$ is changed by $\pm 1$. 
This allows one to determine the relative spin configurations in successive 
layers in the SLs from the obtained sign of the interlayer coefficient. 
We note that this is possible under a strong assumption that the 
structures are morphologically perfect, i.e., with the same, well defined 
$m$ and $n$ values throughout the entire (EuTe)$_m|$(PbTe)$_n$ SL composed 
of several hundreds of periods. From such an analysis it turned out 
that the  spectra for the superlattices with nominally even $m$ and even $n$ 
reveal the preference for the same monolayer spin sequence in successive 
EuTe layers 
($\cdot\cdot\uparrow\downarrow\cdot\cdot\uparrow\downarrow\cdot\cdot$). 
For SLs with odd $m$ and even $n$ neutron diffraction spectra indicated
reversed configuration
($\cdot\cdot\uparrow\downarrow\uparrow\cdot\cdot\downarrow\uparrow
\downarrow\cdot\cdot$). Both these configurations are in agreement 
with the theoretical model prediction.
In contrast, for the samples with $m$ and $n$ both nominally odd, 
the neutron diffraction spectra seem to indicate that the
($\cdot\cdot\cdot\uparrow\downarrow\uparrow\cdot\cdot\cdot\uparrow\downarrow
\uparrow\cdot\cdot\cdot$) 
configuration is preferred, contrary to the theoretical results.

\begin{figure}
\includegraphics[width=3in]{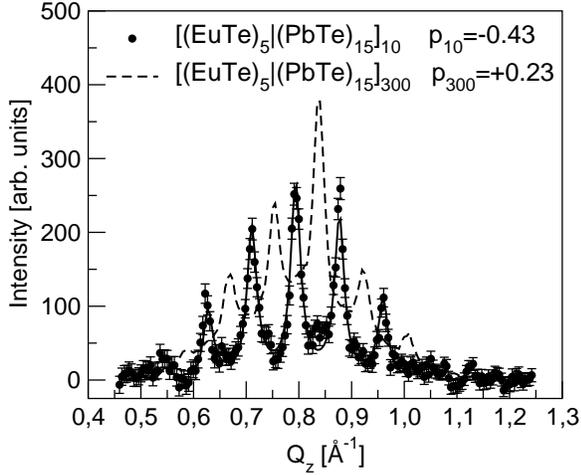}
\caption{\label{t1232} Magnetic diffraction pattern from
highly perfect EuTe/PbTe $(5/15)\times 10$ SL (circulated solid line)
showing the strong interlayer correlation. 
In this case the negative sign of $p$ is in agreement
with the theoretical model predictions. The 
experimental data for the less perfect SL with 300 repetitions (dashed line)
is shown for comparison.}
\end{figure}       
To shed light on this issue, an effort to detect the interlayer exchange 
coupling
in EuTe/PbTe SLs with a smaller number of SL periods, i.e., in SLs with
better controlled $m$ and $n$ values, was undertaken. This task is
not trivial as the intensity of neutron diffraction spectrum 
depends crucially on the number of spins involved. 
From the additional series of
EuTe/PbTe SLs with only 10 periods, one sample indeed showed
SL satellite peaks in positions corresponding to the sign of the coupling 
predicted by the model. In Fig.~\ref{t1232} we present the comparison 
of the spectra of two SLs, both with nominal $m=5$ and $n=15$, and 
with different number of SL periods $N$.
Clearly, the sample with only 10 periods shows the expected negative $p$
value as compared to the opposite sign for the previously measured sample 
with 300
periods (dashed line in Fig.~\ref{t1232}).
This result seems to suggest that in the long process of
MBE growth of the SLs with a large number of repetitions
(typical growth time of several hours) the
preference occurs to form terraces with  even number of monolayers. The reasons
leading to such tendency remain unclear, but one conceivable 
explanation may be  that
the number of monolayers which form a unit cell of the bulk material is
somehow preferred during the long process of layer growth. In the [111] 
direction this corresponds to two cation and two anion sheets. 
Such a mechanism would be valid for both, EuTe and PbTe, constituent materials.
\begin{figure}
\includegraphics[width=3in]{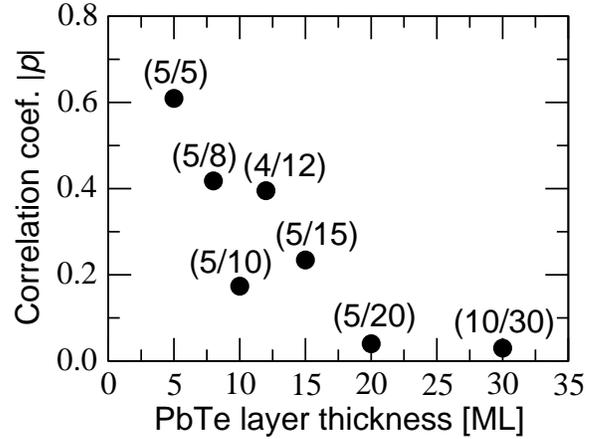}
\caption{\label{pvsn} Dependence of the interlayer correlation parameter 
$p$ on the PbTe spacer thickness in EuTe/PbTe superlattices as determined from
the fitting of the neutron diffraction spectra.}
\end{figure}

The theoretically predicted decrease of the strength of the coupling 
with the nonmagnetic 
spacer thickness is reflected in the decrease of the values of the 
correlation coefficient $p$, as shown in Fig.~\ref{pvsn}. 
A quantitative comparison 
between the experimental and theoretical results is not possible in this 
respect. As far as the range of the interaction is concerned, 
the experimentally observed very long range of the interlayer 
interactions seems to exceed the range predicted by the model. 
The weak correlations still visible in samples with very thick spacers can be 
ascribed to the possible contribution from a residual dipole-dipole 
interaction.

\section{Conclusions}
\label{conclusions}

We have performed an extensive study of the magnetic properties of
AFM EuTe epitaxial layers and of [(EuTe)$_m|$(PbTe)$_n$]$_N$ SLs grown  
by molecular beam epitaxy along the [111] direction.
The structural properties of these samples
were characterized by high resolution x-ray diffraction. The
magnetization and neutron diffraction experiments show that the
magnetic properties of the type II AFM EuTe layers depend sensitively on 
lattice distortions stemming from EuTe/PbTe lattice constant mismatch.
Due to the resulting biaxial strain and finite EuTe layer thickness, 
the FM spin-sheets are all oriented parallel to the (111) growth plane, i.e., 
they form a single $T$-domain.

For a large number of EuTe/PbTe SLs, a systematic study of
the N\'eel temperature dependence on the number of EuTe and PbTe monolayers 
in the SL period was performed. The transition temperature to the AFM 
phase depends on the strain state of the magnetic layers as well as on
their finite thickness.  The observed changes of the N\'eel temperature are
described by the dependence of the exchange parameters on the lattice 
distortions and they follow essentially a mean field behavior. 

From neutron diffraction measurements in an applied
magnetic field parallel to the (111) growth plane, detailed information on the
$S$-- type domain structure  and on the in-plane anisotropy
fields was obtained. The latter ones are considerably higher than
in bulk EuTe. These fields play an important role  in the 
formation of interlayer correlations. Neutron diffraction experiments in 
moderate magnetic fields and magnetization measurements showed that,
irrespective of the number of monolayers in the EuTe layer, no
net magnetic moment is present in the studied SLs. In this sense the 
EuTe/PbTe system constitutes a prototypical example of an 
antiferromagnetic/diamagnetic superlattice.

The most interesting feature of these SLs is a pronounced
interlayer spin correlation between successive EuTe layers revealed by 
magnetic neutron diffraction. The characteristic fingerprint of these
correlations are SL satellite peaks in the vicinity of 
($\frac{1}{2}\frac {1}{2}\frac{1}{2}$) reciprocal lattice point. 
The correlations persist up to PbTe layer thicknesses of about 60 \AA.
Based on kinematical diffraction theory, the formula describing the diffracted
beam intensity as a function of momentum transfer, $Q_z$, has been derived for
a general case of partially correlated SLs. A correlation parameter $p$ 
(obtained by least-square-fitting to the neutron diffraction spectra) was 
found to follow a downward trend with increasing thickness of the nonmagnetic
spacer layer, thus, reflecting the weakening of the interlayer interactions
with the distance between the magnetic layers. 
The  signs of $p$  -- that govern the spin sequences in successive EuTe 
layers -- were compared with the predictions of the theoretical model 
presented in Ref.~\onlinecite{blin}. In this model the
interlayer correlations are mediated by valence-band electrons and are
inferred from the minimization of the total electronic energy of the EuTe/PbTe
system on the spin arrangements in adjacent magnetic layers.
Essentially, the major features of this theoretical model, namely:
\begin{itemize}
\item[(i)]{monotonic decay of the interlayer interactions with the
distance between magnetic layers,}
\item[(ii)]{independence of coupling strength on magnetic layer 
thickness, and }
\item[(iii)]{opposite directions of the spins in the bounding monolayers 
of the two consecutive EuTe layers facing each other across the PbTe spacer}
\end{itemize}
have been confirmed in our neutron and magnetization experiments, although
in order to be able to check experimentally the last issue (especially for
$m-\text{odd}/n-\text{odd}$ SLs) samples with extreme structural perfection 
were necessary.

\appendix*
\section{Neutron diffraction spectrum for AFM superlattices}

In this Appendix we calculate the profiles of 
diffraction spectra for a SL made up of alternating $N$ antiferromagnetic 
layers 
and $N$ nonmagnetic layers, each consisting of $m$ and $n$ atomic monolayers, 
respectively. For simplicity, in the following it is assumed that in both SL
constituent materials the spacing between the monolayers has the 
same value of $d$ and that there are only two possible directions of
spin (magnetization) in each magnetic monolayer (it can be shown,
however, that the results
of this Appendix remain valid  for systems with several in-plane easy axes and
for layers consisting of many $S$-domains  \cite{goldi2}).

We consider three different situations: 
\begin{itemize}
\item[(a)]{{\it perfectly correlated SLs} -- interlayer correlations 
lead to one of the two types of magnetic order, illustrated in the diagrams 
in Section~\ref{theory1}, in the entire SL;} 
\item[(b)] {{\it uncorrelated SLs}; and}
\item[(c)]{{\it partially correlated SLs} -- structures, in which there
is a dominant tendency to form one type of correlations
between the successive magnetic layers, but due to some disruptive
mechanisms a minority of nearest-neighbor layers pairs is aligned in the
opposite way.}
\end{itemize}  

In the standard kinematical theory approach the diffracted wave, 
resulting from magnetic scattering of unpolarized neutrons, 
is obtained by adding up all waves diffracted by individual magnetic atoms. 
This leads to the following equation: 
\begin{equation}
\psi_{\rm diff}\propto f(Q)\sum_{j}\kappa_j \exp (i{\vec Q}\cdot{\vec r}_j)
\label{ap1}
\end{equation}
where ${\vec Q}$ is the scattering vector, $f(Q)$ is the single-atom
magnetic formfactor, ${\vec r}_j$ is the position of the $j^{\rm th}$
atom, and $\kappa_j$ is the magnetic scattering amplitude for a single
atom, equal $+\kappa$ or $-\kappa$ for the ``up'' and ``down'' spin
orientation, respectively. In the symmetric reflection geometry, most 
often used in 
diffraction studies of multilayers, the scattering vector is parallel to the
superlattice axis: ${\vec Q}=(0,0,Q_z)$. The summation over
individual atoms can be then replaced by  a summation over the monolayers.
Since for all atoms located in the $l^{\rm th}$ monolayer
${\vec Q}\cdot {\vec r}_j = Q_z\cdot z = Q_z l d$, the equation
simplifies to:
\begin{equation}
\psi_{\rm diff}\propto f(Q)\sum_{l} {\cal M}_l \exp (iQ_z l d)
\label{ap2}
\end{equation}
where ${\cal M}_l$, the sum of magnetic scattering amplitudes
of all atoms residing in the $l^{\rm th}$ monolayer, is proportional
to the monolayer magnetization.  Taking
advantage of the  SL periodicity, one can separate
this equation into a summation over all monolayers within a SL 
``elementary cell''
-- a bilayer (BL) -- and over the $N$ SL repeats. Thus,
for a bilayer, consisting of $m$ magnetic monolayers (and $n$ nonmagnetic ones,
for which all the ${\cal M}_l = 0$), 
one can define the magnetic structure factor $F_{\rm BL}$ as: 
\begin{equation}
F_{\rm BL}(Q_z) \equiv f(Q)\sum_{\mu=0}^{m-1}{\cal M}_\mu \exp (iQ_z \mu d )
\label{ap4}
\end{equation}
Eq.~(\ref{ap2}) can thus be written in the form:
\begin{equation}
\psi_{\rm diff}\propto F_{\rm BL}(Q_z)\times \sum_{\nu=0}^{N-1} 
\xi_\nu\exp (iQ_zD\nu)
\label{ap3}
\end{equation}
where $D=(m+n)d$ is the SL period. 
The spin configuration in the $\nu^{\rm th}$ magnetic layer with respect 
to the first layer is described 
in Eq.~(\ref{ap3}) by the $\xi_\nu$ coefficient, which takes the 
value $+1$ for  the same and
$-1$ for the opposite magnetization sequences.

The intensity $I(Q_z)$ of the diffracted radiation is given by:
\begin{equation}
|\psi_{\rm diff}|^2 \propto 
\sum_{\alpha=0}^{N-1}\sum_{\beta=0}^{N-1}\xi_\alpha\xi_\beta
\exp[iQ_z(\alpha-\beta)] | F_{\rm BL}(Q_z)|^2
\label{ap5}
\end{equation}
where the structure factor square $|F_{\rm BL}(Q_z)|^2$ can be written as:
\begin{equation}
|F_{\rm BL}(Q_z)|^2 \propto
\left \{ \begin{array}{ll}
\frac{\cos^2(mQ_zd/2)}{\cos^2(Q_zd/2)}&{{\rm for~}m \rm{~odd}}\\
&\\
\frac{\sin^2(mQ_zd/2)}{\cos^2(Q_zd/2)}&{{\rm for~}m \rm{~even}}\\
\end{array} \right .
\label{ap6}
\end{equation}
The structure factor term $| F_{\rm BL}(Q_z)|^2$ has
broad maxima (with weak subsidiaries
on both sides) centered at $Q_z=\frac 12 (2\pi/d)$,
$\frac 32 (2\pi/d),\ldots$, i.e., half-way in between the
reciprocal lattice points corresponding to the basic atomic
structure with periodicity $d$.

Calculating the spectrum profiles $I(Q_z)$ for
SLs with {\it perfect interlayer correlations}
requires putting in Eq.~(\ref{ap5}) $\xi_\nu$  appropriate for the
given type of correlation.
The task reduces then to summing geometric progressions, which yields:
for $\xi_\nu=1$ 
\begin{subequations}
\begin{equation}
I(Q_z) \propto |F_{\rm BL}(Q_z)|^2 \frac{\sin^2(NQ_zD/2)}{\sin^2(Q_zD/2)}
\end{equation}
and for $\xi_\nu=(-1)^{\nu+1}$
\begin{equation}
I(Q_z) \propto |F_{\rm BL}(Q_z)|^2 \times
\left \{ \begin{array}{ll}
\frac{\cos^2(NQ_zD/2)}{\cos^2(Q_zD/2)}&{{\rm for~}N \rm{~odd}}\\
&\\
\frac{\sin^2(NQ_zD/2)}{\cos^2(Q_zD/2)}&{{\rm for~}N \rm{~even}}\\
\end{array} \right .
\end{equation}
\label{ap7}
\end{subequations}

These functions consist of sharp maxima at regular intervals
$\Delta Q_z=2\pi /D$. The intensity of the narrow lines is
``modulated'' by the structure factor, what produces the
characteristic groups of peaks.  It should be noted
that a change from one to another type of interlayer correlations 
causes the narrow line positions to shift by $\frac 12 \Delta Q_z$,
which makes possible to detect such a transition.

In order to analyze {\it uncorrelated} and {\it partially correlated} SLs
the double sum in Eq.~(\ref{ap5}) should be rearranged into sums over
different kinds of layer pairs: namely, the sum of all
same-layer terms ($\alpha=\beta$), the sum of
all terms with $|\alpha-\beta|=1$ (i.e., corresponding
to adjacent magnetic layers),
all terms with $|\alpha-\beta|=2$ (i.e., corresponding
to next-nearest layer pairs), an so on:
\begin{widetext}
\begin{eqnarray}
\sum_{\alpha=0}^{N-1}\sum_{\beta=0}^{N-1}\xi_\alpha\xi_\beta
\exp[iQ_z(\alpha-\beta)]
=N + \sum_{\alpha\ne\beta}\xi_\alpha\xi_\beta\cos[Q_zD(\alpha-\beta)]
\nonumber \nonumber\\
=N + 2\cos(Q_zD)\sum_{\alpha=0}^{N-2}\xi_\alpha \xi_{\alpha+1} +
2\cos(2Q_zD)\sum_{\alpha=0}^{N-3}\xi_\alpha \xi_{\alpha+2}
+\ldots+ 2\cos[(N-1)Q_zD]\xi_0\xi_{N-1}
\label{ap8}
\end{eqnarray}
\end{widetext}

The $\xi_\alpha\xi_\beta$
product for  any pair of layers, labeled $\alpha$ and $\beta$, 
can be thought of as the {\it correlation coefficient} for
this pair. The
number of layer pairs that are $kD$ apart is $N-k$; hence,
the average correlation coefficient for all such pairs in the
SL structure can be written as:
\begin{equation}
p_k \equiv \langle \xi_\alpha\xi_{\alpha+k}\rangle
 =\frac{1}{N-k}\sum_{\alpha=0}^{N-k-1}\xi_\alpha\xi_{\alpha+k}
\label{ap9}
\end{equation}
Using Eqs.~(\ref{ap8}) and (\ref{ap9}) one can write 
the Eq.~(\ref{ap5}) for diffracted intensity in a simple form:
\begin{equation}
I(Q_z)\propto N|F_{\rm BL}|^2\left [1
+2\sum_{k=1}^{N-2}p_k(1-\frac kN)\cos(Q_zDk)\right ]
\label{ap10}
\end{equation}

For a ``perfectly random'' superlattice the correlation
coefficients for all layer pairs vanish on statistical
averaging. Hence, for the {\it uncorrelated} system:
\begin{equation}
I(Q_z)\propto N|F_{\rm BL}(Q_z)|^2
\label{ap11}
\end{equation}
i.e., the diffraction spectrum reproduces the shape of
the structure factor square -- in agreement with the 
expected result for a random system with no coherence between the waves
scattered by individual layers.

The last situation to discuss is  the {\it partially
correlated} superlattice. 
If it is assumed that the only relevant interactions
are between the nearest layers, and there are no long-range
interactions, which introduce coupling between the more distant
layer pairs, then it is straightforward to show that
the correlation coefficient for second-nearest layers
is $p_2=p_1^2$, for third-nearest layers is $p_3=p_1^3$, etc.
In the following we drop the subscript and denote $p_1$ by
$p$.

If the value of $|p|$ is significantly lower than 1, 
the correlation coefficients rapidly decrease, and
only the first few terms in the sum in Eq.~(\ref{ap10}) are relevant.
If, in addition, the number of repeats $N$ in the SL is
large, one can use the approximation $1-k/N \cong 1$, and
obtain:
\begin{equation}
I(Q_z)\propto |F_{\rm BL}(Q_z)|^2\left [1
+2\sum_{k=1}^{\infty}p^k\cos(Q_zDk)\right ]
\label{ap12}
\end{equation} 
By applying the identity \cite{magnus}:
\begin{equation}
1+2\sum_{k=1}^{\infty}p^k\cos(kx)=\frac{1-p^2}{1-2p\cos(x) + p^2}
\nonumber
\label{ap13}
\end{equation}
one obtains the final formula for the spectrum profile:
\begin{equation}
I(Q_z) \propto |F_{\rm BL}(Q_z)|^2\frac{1-p^2}
{1-2p\cos(Q_zD)+p^2}
\label{ap14}
\end{equation}

\begin{acknowledgments}
Work supported by projects: NSF DMR-0204105, FWF, Vienna, Austria,
GME, Vienna, Austria, Austrian Academy of Sciences (APACS),
FENIKS project (EC:G5RD-CT-2001-00535) and the Polish 
State Committee for Scientific Research grant PBZ-KBN-044/P03/2001.
The collaboration with J. K. Furdyna and M. S. Dresselhaus and their 
contribution to the initial stage of these studies is acknowledged.
\end{acknowledgments}

\end{document}